\documentclass[twocolumn,english,aps,pra,showpacs]{revtex4}
\usepackage[T1]{fontenc}
\usepackage[latin1]{inputenc}
\usepackage{amsmath}
\usepackage{amssymb}
\usepackage{graphicx}
\usepackage{esint}

\makeatletter
\@ifundefined{textcolor}{}
{%
 \definecolor{BLACK}{gray}{0}
 \definecolor{WHITE}{gray}{1}
 \definecolor{RED}{rgb}{1,0,0}
 \definecolor{GREEN}{rgb}{0,1,0}
 \definecolor{BLUE}{rgb}{0,0,1}
 \definecolor{CYAN}{cmyk}{1,0,0,0}
 \definecolor{MAGENTA}{cmyk}{0,1,0,0}
 \definecolor{YELLOW}{cmyk}{0,0,1,0}
}

\usepackage{babel}

\usepackage{babel}

\usepackage{babel}

\makeatother

\usepackage{babel}
\begin{document}

\title{Impurity probe of topological superfluid in one-dimensional spin-orbit
coupled atomic Fermi gases}

\author{Xia-Ji Liu$^{1}$ }

\email{xiajiliu@swin.edu.au}

\selectlanguage{english}%

\affiliation{$^{1}$ARC Centre of Excellence for Quantum-Atom Optics, Centre for
Atom Optics and Ultrafast Spectroscopy, Swinburne University of Technology,
Melbourne 3122, Australia}

\date{\today}
\begin{abstract}
We investigate theoretically non-magnetic impurity scattering in a
one-dimensional atomic topological superfluid in harmonic traps, by
solving self-consistently the microscopic Bogoliubov-de Gennes equation.
In sharp contrast to topologically trivial Bardeen-Cooper-Schrieffer
\textit{s}-wave superfluid, topological superfluid can host a mid-gap
state that is bound to localized non-magnetic impurity. For strong
impurity scattering, the bound state becomes universal, with nearly
zero energy and a wave-function that closely follows the symmetry
of that of Majorana fermions. We propose that the observation of such
a universal bound state could be a useful evidence for characterizing
the topolgoical nature of topological superfluids. Our prediction
is applicable to an ultracold resonantly-interacting Fermi gas of
$^{40}$K atoms with spin-orbit coupling confined in a two-dimensional
optical lattice. 
\end{abstract}

\pacs{03.75.Ss, 71.10.Pm, 03.65.Vf, 03.67.Lx}

\maketitle

\section{Introduction}

Impurity scattering plays an important role in understanding the quantum
state of hosting systems \cite{Balatsky2006}. This is particularly
significant in solid state systems, where impurity scattering and
disorder are intrinsic. In superconductors, the study of impurity
effects has the potential to uncover the nature and origin of the
superconducting state \cite{Mackenzie1998}. In strongly correlated
electronic systems near quantum critical points, where several types
of ordering compete in a delicate balance, the study of impurity scatterings
has the power to underpin in favor of one of the orders \cite{Millis2003}.
In this work, we aim to investigate theoretically impurity scattering
in one-dimensional (1D) topological superfluids. We show that an impurity-induced
bound state will provide a sensitive probe for the topological order
in such systems.

Topological superfluid is a novel state of quantum matter \cite{Qi2011},
which is gapped in the bulk, but hosts non-trivial zero-energy surface
states - the called Majorana fermions \cite{Majorana1937,Wilczek2009}
- near its boundary. It has attracted great attentions in recent years
because of its potential application in topological quantum computation
and quantum information \cite{Kitaev2006,Nayak2008}. The realization
of topological superfluids and the manipulation of Majorana fermions
are currently the most hot research topic in a variety fields of physics,
ranging from condensed matter physics to ultracold atomic systems.
Till now, indirect evidence of the existence of topological superfluids
in hybrid superconductor-semiconductor InSb or InAs nanowires has
been reported \cite{Mourik2012,Rokhinson2012,Das2012}. Theoretical
schemes of processing topological quantum information in such nanowire
devices have also been proposed \cite{Fu2008,Alicea2011}.

Our investigation of impurity scattering in 1D topological superfluids
is strongly motivated by the rapid experimental progress \cite{Mourik2012,Rokhinson2012,Das2012}.
On one hand, impurity scattering is un-avoidable in InSb or InAs nanowires.
A realistic simulation of impurity scattering may therefore be useful
for future solid-state experiments. On the other hand, we anticipate
that impurity may induce new exotic bound state, thus providing a
clear local probe of the topological nature of the systems that we
consider.

In this paper, we use a 1D spin-orbit coupled atomic Fermi gas to
model 1D topological superfluids \cite{Jiang2011,Liu2012,Wei2012},
instead of considering nanowire devices used in solid-state \cite{Mourik2012,Rokhinson2012,Das2012}.
This is because we have unprecedented controllability with ultracold
atomic gases \cite{Bloch2008}. By using magnetic Feshbach resonances,
the interatomic interactions can be precisely tuned \cite{Chin2010}.
Using the technique of optical lattices, artificial 1D and 2D environments
can be easily created \cite{Hu2007,Liao2010}. The spin-orbit coupling,
which is the necessary ingredient of a realistic topological superfluid,
can also be engineered with arbitrary strength \cite{Wang2012,Cheuk2012}.
Thus, ultracold spin-orbit coupled atomic Fermi gas is arguably the
best candidate to simulate the desired topological superfluids. Furthermore,
even though cold atom systems are intrinsically clean, individual
impurities can be realized using off-resonant dimple laser light or
another species of atoms or ions \cite{OurImpurity}. The disorder
effect of many randomly distributed impurities can also be created
by employing quasiperiodic bichromatic lattices or laser speckles
\cite{Palencia2010}.

We investigate the impurity effect in 1D spin-orbit coupled atomic
Fermi gas of $^{40}$K atoms by solving self-consistently the microscopic
Bogoliubov-de Gennes (BdG) equation, with realistic experimental parameters.
We observe the existence of mid-gap state that is bound to localized
non-magnetic impurity. For strong impurity scattering, the bound state
tends to be universal, with nearly zero energy and a wave-function
that closely follows the symmetry of that of Majorana fermions. This
feature is clearly absent in topologically trivial superfluids. Therefore,
we argue that the observation of such a universal bound state would
be a useful evidence for characterizing the topological nature of
topological superfluids. We note that, mid-gap bound state induced
by non-magnetic impurity has also been predicted in 1D spin-orbit
coupled superconductors, by using non-self-consistent \textit{T}-matrix
theory \cite{Sau2012}. The effect of magnetic impurity in 2D spin-orbit
coupled Fermi gases has also been studied analytically using \textit{T}-matrix
formalism \cite{Yan2012}.

Our paper is arranged as follows. In the next section (Sec. II), we
introduce briefly the model Hamiltonian and the solution of BdG equations,
and then present a phase diagram for a given set of experimental parameters.
In Sec. III, we study non-magnetic impurity scatterings and show the
emergence of universal bound state in the strong scattering limit.
The properties of such a universal bound state are analyzed in greater
detail. To better simulate the realistic experimental setup, we also
consider an extended impurity with gaussian-shape scattering potential.
Finally, we summarize in Sec. IV. The detailed numerical procedure
of solving BdG equations is listed in the Appendix A, together with
a careful check on numerical accuracy.

\section{Model Hamiltonian and BdG equations}

The framework of our theoretical approach has been briefly described
in our previous work \cite{Liu2012}. Here, we emphasize on the experimental
origin of the model Hamiltonian and generalize the theoretical approach
to include a classical non-magnetic impurity. A detailed discussion
on the numerical procedure is given in the Appendix A.

\subsection{1D spin-orbit coupled Fermi gas}

Let us consider a spin-orbit-coupled Fermi gas of $^{40}$K atoms
in harmonic traps, realized recently at Shanxi University \cite{Wang2012}.
We assume additional confinement due to a very deep 2D optical lattice
in the transverse $y-z$ plane, which restricts the motion of atoms
to the $x$-axis. The spin-orbit coupling is created by two counter
propagating Raman laser beams that couple the two spin states of the
system along the $x$-axis \cite{Wang2012}. Near the Feshbach resonance
$B_{0}\simeq202.20$ G, the quasi-1D Fermi system may be described
by a single-channel model Hamiltonian $H=H_{0}+H_{int}$, where 
\begin{eqnarray}
H_{0} & = & \sum_{\sigma=\uparrow,\downarrow}\int dx\Psi_{\sigma}^{\dagger}\left[-\frac{\hbar^{2}}{2m}\frac{\partial^{2}}{\partial x^{2}}-\mu+V_{T}\right]\Psi_{\sigma}\left(x\right)\nonumber \\
 &  & -\frac{\Omega_{R}}{2}\int dx\left[\Psi_{\uparrow}^{\dagger}\left(x\right)e^{i2k_{R}x}\Psi_{\downarrow}\left(x\right)+\text{H.c.}\right]\label{bareHami1}
\end{eqnarray}
 is the single-particle Hamiltonian in the presence of Raman process
and 
\begin{equation}
H_{int}=g_{1D}\int dx\Psi_{\uparrow}^{\dagger}\left(x\right)\Psi_{\downarrow}^{\dagger}\left(x\right)\Psi_{\downarrow}\left(x\right)\Psi_{\uparrow}\left(x\right)
\end{equation}
 is the interaction Hamiltonian describing the contact interaction
between two spin states. Here, the pseudospins $\sigma=\uparrow,\downarrow$
denote the two hyperfine states, and $\Psi_{\sigma}\left(x\right)$
is the Fermi field operator that annihilates an atom with mass $m$
at position $x$ in the spin $\sigma$ state. The chemical potential
$\mu$ is determined by the total number of atoms $N$ in the system.
For the two-photon Raman process, $\Omega_{R}$ is the coupling strength
of Raman beams, $k_{R}$ $=2\pi/\lambda_{R}$ is determined by the
wave length $\lambda_{R}$ of two lasers and therefore $2\hbar k_{R}$
is the momentum transfer during the process. The trapping potential
$V_{T}\left(x\right)\equiv m\omega^{2}x^{2}/2$ refers to the harmonic
trap with an oscillation frequency $\omega=\omega_{x}$ in the axial
direction. In such a quasi-one dimensional geometry, it is shown by
Bergeman et al. \cite{Bergeman2003} that the scattering properties
of the atoms can be well described using a contact potential $g_{1D}\delta(x)$,
where the 1D effective coupling constant $g_{1D}<0$ may be expressed
through the 3D scattering length $a_{3D}$,

\begin{equation}
g_{1D}=\frac{2\hbar^{2}a_{3D}}{ma_{\perp}^{2}}\frac{1}{\left(1-{\cal A}a_{3D}/a_{\perp}\right)},
\end{equation}
 where $a_{\perp}\equiv\sqrt{\hbar/(m\omega_{\perp})}$ is the characteristic
oscillator length in the transverse axis, for a given transverse trapping
frequency $\omega_{\perp}$ set by the deep 2D optical lattice. The
constant ${\cal A}=-\zeta(1/2)/\sqrt{2}\simeq1.0326$ is responsible
for the confinement induced Feshbach resonance \cite{Bergeman2003},
which changes the scattering properties dramatically when the 3D scattering
length is comparable to the transverse oscillator length. It is also
convenient to express $g_{1D}$ in terms of an effective1D scattering
length, $g_{1D}=-2\hbar^{2}/\left(ma_{1D}\right)$, where $a_{1D}=-(a_{\perp}^{2}/a_{3D})(1-{\cal A}a_{3D}/a_{\perp})>0$.
The interatomic interaction can then be described by a dimensionless
interaction parameter $\gamma\equiv a/[\pi\sqrt{N}a_{1D}]$, where
$a\equiv\sqrt{\hbar/(m\omega)}$ is the oscillator length in the $x$-axis.
Near the Feshbach resonance, the typical value of the interaction
parameter $\gamma$ is about $5$ \cite{Liao2010,Liu2007,Liu2008}.

To illustrate how the spin-orbit coupling is induced by the two-photon
Raman process, it is useful to remove the spatial dependence of the
Raman coupling term, by taking the following local gauge transformation,
\begin{eqnarray}
\Psi_{\uparrow}\left(x\right) & = & e^{+ik_{R}x}\tilde{\psi}_{\uparrow}\left(x\right),\\
\Psi_{\downarrow}\left(x\right) & = & e^{-ik_{R}x}\tilde{\psi}_{\downarrow}\left(x\right).
\end{eqnarray}
 Using the new field operators $\tilde{\psi}_{\uparrow}\left(x\right)$
and $\tilde{\psi}_{\downarrow}\left(x\right)$, we can recast the
single-particle Hamiltonian as 
\begin{eqnarray}
H_{0} & = & \int dx\left[\tilde{\psi}_{\uparrow}^{\dagger}\left(x\right),\tilde{\psi}_{\downarrow}^{\dagger}\left(x\right)\right]{\cal H}_{0}\left[\begin{array}{c}
\tilde{\psi}_{\uparrow}\left(x\right)\\
\tilde{\psi}_{\downarrow}\left(x\right)
\end{array}\right],\\
{\cal H}_{0} & = & -\frac{\hbar^{2}}{2m}\frac{\partial^{2}}{\partial x^{2}}-\mu+V_{T}\left(x\right)-h\sigma_{x}+\lambda\hat{k}_{x}\sigma_{z},
\end{eqnarray}
 where we have absorbed a constant energy shift $E_{R}\equiv\hbar^{2}k_{R}^{2}/(2m)$
(the recoil energy) in the chemical potential $\mu$, and have defined
the momentum operator $\hat{k}_{x}\equiv-i\partial/\partial x$, the
spin-orbit coupling constant $\lambda\equiv\hbar^{2}k_{R}/m$ and
an effective Zeeman field $h\equiv\Omega_{R}/2$. $\sigma_{x}$ and
$\sigma_{z}$ are Pauli's matrices. The spin-orbit coupling in the
Hamiltonian ${\cal H}_{0}$ can be regarded as an equal-weight combination
of Rashba and Dresselhaus spin-orbit coupling (i.e., $\lambda\hat{k}_{x}\sigma_{y}$).
This is evident after we take the second local gauge transformation,
\begin{eqnarray}
\tilde{\psi}_{\uparrow}\left(x\right) & = & \frac{1}{\sqrt{2}}\left[\psi_{\uparrow}\left(x\right)-i\psi_{\downarrow}\left(x\right)\right],\\
\tilde{\psi}_{\downarrow}\left(x\right) & = & \frac{1}{\sqrt{2}}\left[\psi_{\uparrow}\left(x\right)+i\psi_{\downarrow}\left(x\right)\right],
\end{eqnarray}
 with which the single-particle Hamiltonian becomes, 
\begin{eqnarray}
H_{0} & = & \int dx\left[\psi_{\uparrow}^{\dagger}\left(x\right),\psi_{\downarrow}^{\dagger}\left(x\right)\right]{\cal H}_{0}\left[\begin{array}{c}
\psi_{\uparrow}\left(x\right)\\
\psi_{\downarrow}\left(x\right)
\end{array}\right],\\
{\cal H}_{0} & = & -\frac{\hbar^{2}}{2m}\frac{\partial^{2}}{\partial x^{2}}+V_{T}\left(x\right)-\mu-h\sigma_{z}+\lambda\hat{k}_{x}\sigma_{y}.
\end{eqnarray}
 The form of the interaction Hamiltonian is invariant after two gauge
transformations, i.e., 
\begin{equation}
H_{int}=g_{1D}\int dx\psi_{\uparrow}^{\dagger}\left(x\right)\psi_{\downarrow}^{\dagger}\left(x\right)\psi_{\downarrow}\left(x\right)\psi_{\uparrow}\left(x\right).
\end{equation}
 We note that the operator of total density $\hat{n}(x)\equiv\sum_{\sigma}\Psi_{\sigma}^{\dagger}\left(x\right)\Psi_{\sigma}\left(x\right)=\sum_{\sigma}\psi_{\sigma}^{\dagger}\left(x\right)\psi_{\sigma}\left(x\right)$
is also invariant in the gauge transformation.

\subsection{Impurity scattering Hamiltonian}

Now we add the non-magnetic impurity scattering term, 
\begin{equation}
H_{imp}=\int dxV_{imp}\left(x\right)\sum_{\sigma}\psi_{\sigma}^{\dagger}\left(x\right)\psi_{\sigma}\left(x\right),
\end{equation}
 to the total Hamiltonian. The non-magnetic scattering can be realized
experimentally by using an off-resonant dimple laser light. We consider
either a localized scattering potential at position $x_{0}$, 
\begin{equation}
V_{imp}\left(x\right)=V_{imp}\delta\left(x-x_{0}\right),
\end{equation}
 or an extend potential with a width $d$ in the gaussian line-shape,
\begin{equation}
V_{imp}\left(x\right)=\frac{V_{imp}}{\sqrt{2\pi}d}\exp[-\frac{\left(x-x_{0}\right)^{2}}{2d^{2}}].
\end{equation}
 The strength of the impurity scattering is given by $V_{imp}$. In
the narrow width limit $d\rightarrow0$, the gaussian potential returns
back to the delta-like potential. We may place the impurity at arbitrary
position, as long as the Fermi system is locally in the topological
superfluid state. To be concrete, we shall set $x_{0}=0$.

We may also consider a magnetic impurity scattering in the form, $H_{imp}=\int dxV_{imp}(x)[\psi_{\uparrow}^{\dagger}\left(x\right)\psi_{\uparrow}\left(x\right)-\psi_{\downarrow}^{\dagger}\left(x\right)\psi_{\downarrow}\left(x\right)]$.
However, it is of theoretical interest only. The field operator of
density difference is not invariant in the second local gauge transformation.
Thus, experimentally the magnetic impurity scattering potential is
more difficult to realize.

\subsection{Bogoliubov-de Gennes equation}

We use the standard mean-field theory to solve the model Hamiltonian.
By introducing a real order parameter $\Delta\left(x\right)\equiv-g_{1D}\left\langle \psi_{\downarrow}\left(x\right)\psi_{\uparrow}\left(x\right)\right\rangle $,
the interaction Hamiltonian is decoupled as, 
\begin{equation}
H_{int}\simeq-\int dx\left[\Delta\left(x\right)\psi_{\uparrow}^{\dagger}\psi_{\downarrow}^{\dagger}\left(x\right)+\text{H.c.}+\frac{\left|\Delta\left(x\right)\right|^{2}}{g_{1D}}\right].
\end{equation}
 It is then convenient to introduce a Nambu spinor ${\bf \mathbf{\boldsymbol{\psi}}}(x)\equiv[\psi_{\uparrow}\left(x\right),\psi_{\downarrow}\left(x\right),\psi_{\uparrow}^{\dagger}\left(x\right),\psi_{\downarrow}^{\dagger}\left(x\right)]^{T}$
and rewrite the mean-field Hamiltonian in a compact form, 
\begin{equation}
H_{mf}=\frac{1}{2}\int dx{\bf \boldsymbol{\psi}}^{\dagger}{\cal H}_{BdG}{\bf \boldsymbol{\psi}}(x)+\text{Tr}{\cal H}_{S}-\int dx\frac{\left|\Delta\left(x\right)\right|^{2}}{g_{1D}},\label{mf_Hami}
\end{equation}
 where

\begin{equation}
{\cal H}_{BdG}=\left[\begin{array}{cccc}
{\cal H}_{S}-h & -\lambda\partial/\partial x & 0 & -\Delta(x)\\
\lambda\partial/\partial x & {\cal H}_{S}+h & \Delta(x) & 0\\
0 & \Delta^{*}(x) & -{\cal H}_{S}+h & \lambda\partial/\partial x\\
-\Delta^{*}(x) & 0 & -\lambda\partial/\partial x & -{\cal H}_{S}-h
\end{array}\right]\label{BdG_Hami}
\end{equation}
 and 
\begin{equation}
{\cal H}_{S}(x)\equiv-\frac{\hbar^{2}}{2m}\frac{\partial^{2}}{\partial x^{2}}-\mu+\frac{m}{2}\omega^{2}x^{2}+V_{imp}(x).
\end{equation}
 The term Tr${\cal H}_{S}$ in $H_{mf}$ results from the anti-commutativity
of Fermi field operators.

The mean-field Hamiltonian Eq. (\ref{mf_Hami}) can be diagonalized
by the standard Bogoliubov transformation. By defining the field operators
$\alpha_{\eta}$ for Bogoliubov quasiparticles, 
\begin{equation}
\alpha_{\eta}=\int dx\sum_{\sigma}\left[u_{\sigma\eta}\left(x\right)\psi_{\sigma}\left(x\right)+\nu_{\sigma\eta}\left(x\right)\psi_{\sigma}^{\dagger}\left(x\right)\right],
\end{equation}
 we obtain that, 
\begin{equation}
H_{mf}=\frac{1}{2}\sum_{\eta}E_{\eta}\alpha_{\eta}^{\dagger}\alpha_{\eta}+\text{Tr}{\cal H}_{S}-\int dx\frac{\left|\Delta\left(x\right)\right|^{2}}{g_{1D}},
\end{equation}
 Here, $\Phi_{\eta}(x)\equiv[u_{\uparrow\eta}\left(x\right),u_{\downarrow\eta}\left(x\right),v_{\uparrow\eta}\left(x\right),v_{\downarrow\eta}\left(x\right)]^{T}$
and $E_{\eta}$ are respectively the wave-function and energy of Bogoliubov
quasiparticles, satisfying the BdG equation, 
\begin{equation}
{\cal H}_{BdG}\Phi_{\eta}\left(x\right)=E_{\eta}\Phi_{\eta}\left(x\right).\label{bdgeq}
\end{equation}
 The BdG Hamiltonian Eq. (\ref{BdG_Hami}) includes the pairing gap
function $\Delta\left(x\right)$ that should be determined self-consistently.
For this purpose, we take the inverse Bogoliubov transformation and
obtain 
\begin{equation}
\psi_{\sigma}\left(x\right)=\sum_{\eta}\left[u_{\sigma\eta}\left(x\right)\alpha_{\eta}+\nu_{\sigma\eta}^{*}\left(x\right)\alpha_{\eta}^{\dagger}\right].
\end{equation}
 The gap function $\Delta\left(x\right)$ is then given by, 
\begin{eqnarray}
\Delta(x) & = & -\frac{g_{1D}}{2}\sum_{\eta}\left[u_{\uparrow\eta}\left(x\right)v_{\downarrow\eta}^{*}\left(x\right)f\left(E_{\eta}\right)\right.\nonumber \\
 &  & \left.+u_{\downarrow\eta}\left(x\right)v_{\uparrow\eta}^{*}\left(x\right)f\left(-E_{\eta}\right)\right],\label{gapeq}
\end{eqnarray}
 where $f\left(E\right)\equiv1/[e^{E/k_{B}T}+1]$ is the Fermi distribution
function at temperature $T$. Accordingly, the total density take
the form, 
\begin{equation}
n\left(x\right)=\frac{1}{2}\sum_{\sigma\eta}\left[\left|u_{\sigma\eta}\left(x\right)\right|^{2}f\left(E_{\eta}\right)+\left|v_{\sigma\eta}\left(x\right)\right|^{2}f\left(-E_{\eta}\right)\right].\label{numeq}
\end{equation}
 The chemical potential $\mu$ can be determined using the number
equation, $N=\int dxn\left(x\right)$.

It is important to note that, the use of Nambu spinor representation
enlarges the Hilbert space of the system. As a result, there is an
intrinsic particle-hole symmetry in the Bogoliubov solutions \cite{Liu2007,Liu2008}:
for any ``particle'' solution with the wave-function $\Phi_{\eta}^{(p)}(x)=[u_{\uparrow\eta}\left(x\right),u_{\downarrow\eta}\left(x\right),v_{\uparrow\eta}\left(x\right),v_{\downarrow\eta}\left(x\right)]^{T}$
and energy $E_{\eta}^{(p)}\geq0$, we can always find another partner
``hole'' solution with the wave-function $\Phi_{\eta}^{(h)}(x)=[v_{\uparrow\eta}^{*}\left(x\right),v_{\downarrow\eta}^{*}\left(x\right),u_{\uparrow\eta}^{*}\left(x\right),u_{\downarrow\eta}^{*}\left(x\right)]^{T}$
and energy $E_{\eta}^{(h)}=-E_{\eta}^{(p)}\leq0$. In general, these
two solutions correspond to the same physical state. To remove this
redundancy, we have added an extra factor of 1/2 in the expressions
for pairing gap function Eq. (\ref{gapeq}) and total density Eq.
(\ref{numeq}).

The Bogoliubov equation Eq. (\ref{BdG_Hami}) can be solved iteratively
with Eqs. (\ref{gapeq}) and (\ref{numeq}) by using a basis expansion
method, together with a hybrid strategy that takes care of the high-lying
energy states \cite{Liu2012,Liu2007,Liu2008}. A detailed discussion
on the numerical procedure and a self-consistent check on the numerical
accuracy are outlined in the Appendix A.

\subsection{Phase diagram in the absence of impurity}

In our previous study \cite{Liu2012}, we have discussed the phase
diagram of a weakly interacting spin-orbit coupled Fermi gas, with
an interaction parameter $\gamma=\pi/2\simeq1.6$. The real experiment,
however, would be carried out near Feshbach resonances, where the
typical interaction parameter is $\gamma=3\sim5$ \cite{Liao2010,Liu2007,Liu2008}.
In this work, we take a realistic interaction parameter $\gamma=\pi\simeq3.2$,
despite the fact that our mean-field treatment would become less accurate.
We consider a Fermi gas of $N=100$ atoms in a single tube formed
by a tight 2D optical lattice, and take the Thomas-Fermi energy $E_{F}=k_{B}T_{F}=(N/2)\hbar\omega$
and Thomas-Fermi radius $x_{F}=\sqrt{N}a$ as the units for energy
and length, respectively. For the spin-orbit coupling, we use a dimensionless
parameter $\lambda k_{F}/E_{F}=1$, where $k_{F}=\sqrt{2mE_{F}}$
is the Thomas-Fermi wavevector.

\begin{figure}[t]
\begin{centering}
\includegraphics[clip,width=0.48\textwidth]{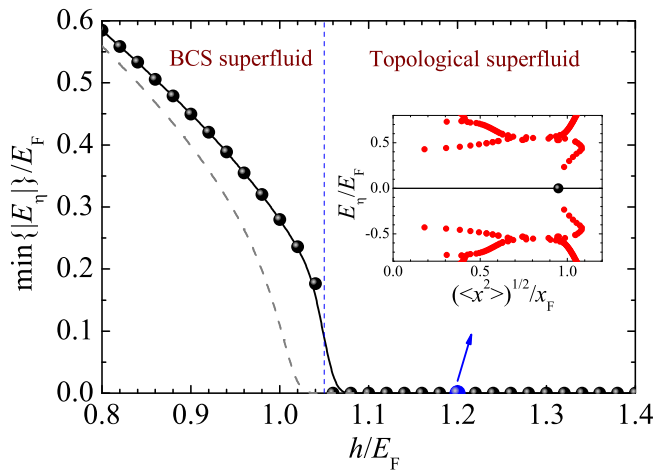} 
\par\end{centering}

\caption{(color online) Phase diagram at $T=0$ (solid line) and $T=0.3T_{F}$
(gray dashed line), determined from the behavior of the lowest energy
in quasiparticle spectrum. With increasing the effective Zeeman field,
the Fermi cloud changes from a standard BCS superfluid to a topologically
non-trivial superfluid. The phase transition point is slightly affected
by finite temperature. The inset shows the energy spectrum at $h=1.2E_{F}$
as a function of the position of quasiparticles. A zero-energy quasiparticle
(i.e., Majorana fermion) at the trap edge has been highlighted by
a big dark circle. Here, we characterize approximately the position
of a quasiparticle by using its wave-function: $\left\langle x^{2}\right\rangle =\int dxx^{2}\sum_{\sigma}[u_{\sigma}^{2}\left(x\right)+\nu_{\sigma}^{2}\left(x\right)]$. }

\label{fig1} 
\end{figure}

Fig. 1 presents the phase diagram at these parameters and at two temperatures
$T=0$ and $T=0.3T_{F}$, showing the well-known topological phase
transition at a critical effective Zeeman field $h_{c}\simeq1.05E_{F}$.
The different phase is characterized by the lowest energy of Bogoliubov
quasiparticles, $\min\{\left|E_{\eta}\right|\}$. At a small Zeeman
field $h<h_{c}$, the system is a standard Bardeen-Cooper-Schrieffer
(BCS) superfluid, with a fully gapped quasiparticle energy spectrum
(i.e., $\min\{\left|E_{\eta}\right|\}>0$). Once $h>h_{c}$, however,
topologically non-trivial phase emerges. Though the quasiparticle
energy spectrum is still gapped in the bulk, gapless excitations -
Majorana fermions - appear at the edges %
\footnote{The energy of the gapless excitations is not precisely zero, due to
the finite size of the system. It scales exponentially with the cloud
size. Typically, it is about $10^{-10}E_{F}$.%
}, leading to an exponentially small lowest energy in the spectrum.
This is fairly evident in the inset, where we plot the energy spectrum
as a function of the position of quasiparticles.

\begin{figure}[htp]
\begin{centering}
\includegraphics[clip,width=0.48\textwidth]{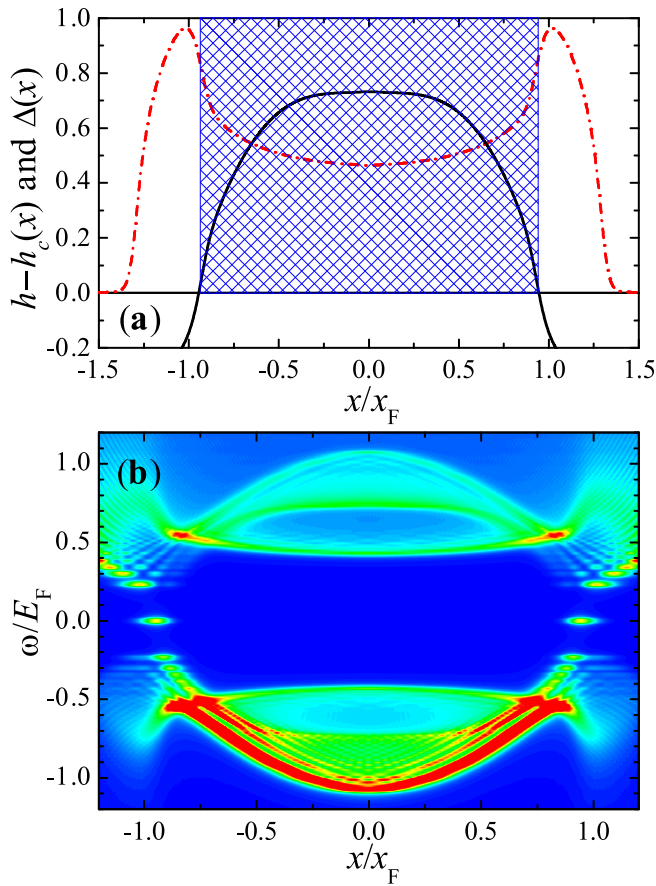} 
\par\end{centering}

\caption{(color online) (a) The pairing gap distribution function $\Delta(x)$
(dot-dashed line) and the criterion for a local topological phase
$h>h_{c}(x)$ (solid line) at $h=1.2E_{F}$. The shaded cross-hatching
highlights the topologically non-trivial area. (b) The linear contour
plot of the local density of state at $h=1.2E_{F}$. At each trap
edge, a series of edge states, including the zero-energy Majorana
fermion mode, are clearly visible. }

\label{fig2} 
\end{figure}

To determine the critical Zeeman field $h_{c}$, we note that for
a homogeneous spin-orbit coupled Fermi gas, it is given by \cite{Lutchyn2010,Oreg2010}
\begin{equation}
h_{c}=\sqrt{\mu^{2}+\Delta^{2}}.\label{eq:hc}
\end{equation}
 In harmonic traps as we consider here, the critical Zeeman field
becomes position dependent. The local critical Zeeman field, calculated
using $h_{c}(x)=\sqrt{\mu^{2}(x)+\Delta^{2}(x)}$, with the local
chemical potential $\mu(x)=\mu-m\omega^{2}x^{2}/2$ and the local
pairing gap $\Delta(x)$, increases monotonically towards the trap
edge %
\footnote{For weak attractive interactions, the local critical Zeeman field
may decrease towards the trap edge. As a result, the topological phase
appears first at the trap edge, leading to a phase-separation phase
consisting of a topological superfluid at the edge and a BCS superfluid
at the center. See, for example, Ref. \cite{Liu2012} for more details.%
}. The Fermi cloud at position $x$ will locally be in a topological
state if the Zeeman field $h>h_{c}(x)$. For the parameters given
in the above, this first happens at $h\simeq1.05E_{F}$, for which
the local phase at the trap center ($x=0$) starts to become topologically
non-trivial. In Fig. 2(a), we show the local pairing gap $\Delta(x)$
and the criterion for a local topological state, $h>h_{c}(x)$, at
the Zeeman field $h=1.2E_{F}$. At this field, the topological area
is extended to the edge of the trap, as highlighted by a shaded cross-hatching.
The appearance of Majorana fermion modes may be probed by measuring
the local density of state through spatially resolved radio-frequency
(rf) spectroscopy \cite{Liu2012,OurImpurity}. In Fig. 2(b), we present
the local density of state at $h=1.2E_{F}$, 
\begin{eqnarray}
\rho\left(x,\omega\right) & = & \frac{1}{2}\sum_{\sigma\eta}\left[\left|u_{\sigma\eta}\left(x\right)\right|^{2}\delta\left(\omega-E_{\eta}\right)\right.\nonumber \\
 &  & +\left.\left|v_{\sigma\eta}\left(x\right)\right|^{2}\delta\left(\omega+E_{\eta}\right)\right].
\end{eqnarray}
 At each of the two trap edges, we observe a series of edge states,
whose dispersion relation is approximately given by \cite{Wei2012},
$E_{n}=\sqrt{n}\Delta E$, where $n=0,1,2,\cdots$ is a non-negative
integer and $\Delta E$ is a characteristic energy scale set by the
trapping frequency and recoil energy. The Majorana fermion modes with
zero energy $E_{n}=0$ are clearly visible.

\section{Universal impurity-induced bound state}

We are now ready to investigate how Bogoliubov quasiparticles are
affected by a non-magnetic impurity. Hereafter, we focus on the topological
state at $h=1.2E_{F}$. For a topologically trivial state at $h<1.05E_{F}$,
we have checked numerically that quasiparticles are essentially not
affected by the non-magnetic impurity scattering. This is in accord
with the well-known Anderson's theorem that potential scattering impurities
are not pair-breakers in \textit{s}-wave superconductors \cite{Balatsky2006,Anderson1959}.

\subsection{Impurity-induced mid-gap state}

In Fig. 3, we report the density profile and pairing gap distribution
in the presence of a strong non-magnetic impurity with scattering
potential strength, $V_{imp}=-0.30x_{F}E_{F}$. Both of them are completely
depleted at the impurity site $x=0$. Accordingly, we observe the
appearance of a new mid-gap state that is bound to the impurity, as
shown in the inset for the spatial distribution of Bogliubov quasiparticles.
This is clearly seen when we compare the quasiparticle spectrum without
and with the non-magnetic impurity, i.e., the inset in Fig. 1 and
Fig. 3, respectively. Away from the impurity site, the distribution
of Bogoliubov quasiparticles is also disturbed by the impurity. However,
the series of edge states at the trap edge seems to be very robust
against the impurity scattering.

\begin{figure}[htp]
\begin{centering}
\includegraphics[clip,width=0.48\textwidth]{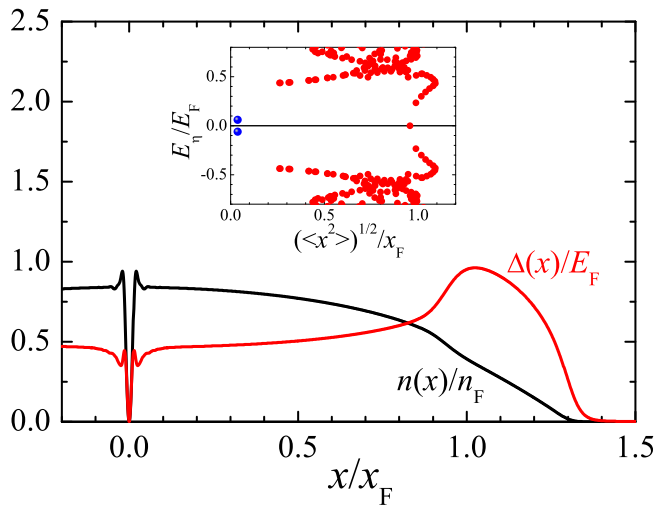} 
\par\end{centering}

\caption{(color online) Density profile and pairing gap distribution in the
presence of a strong attractive non-magnetic impurity with scattering
potential strength, $V_{imp}=-0.30x_{F}E_{F}$. The inset shows the
spatial distribution of Bogoliubov quasiparticle energy spectrum.
The mid-gap bound state near the impurity site $x=0$ is highlighted
by big blue circles. }

\label{fig3} 
\end{figure}

\begin{figure}[htp]
\begin{centering}
\includegraphics[clip,width=0.48\textwidth]{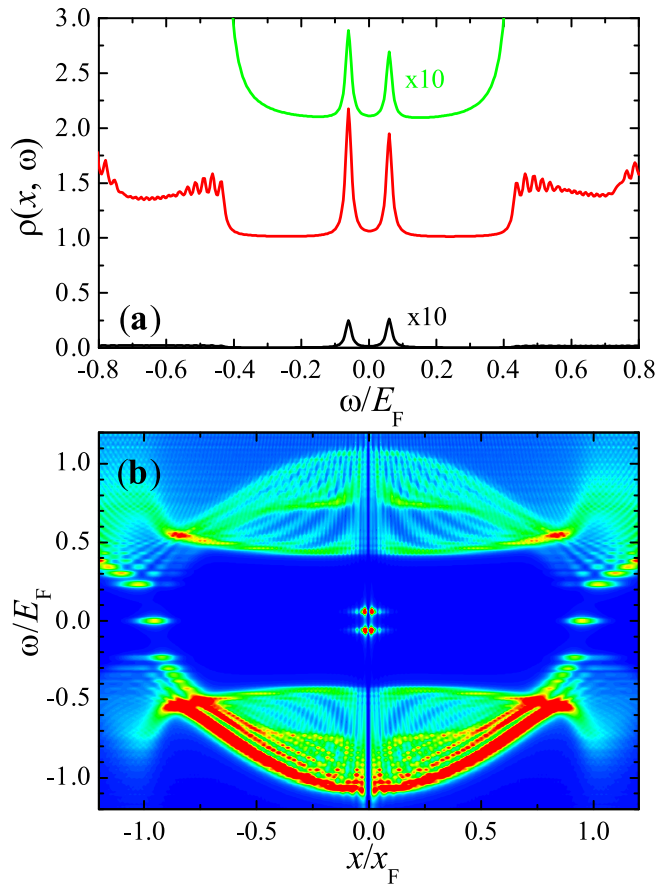} 
\par\end{centering}

\caption{(color online) (a) Density of state for a topological superfluid ($h=1.2E_{F}$),
at $x=0$, $0.05x_{F}$, and $0.1x_{F}$ (from bottom to top). For
better illustration, the curves have been off-set. The magnitude of
the local density of state at $x=0$ and $0.1x_{F}$ has been enlarged
by a factor of $10$. (b) Linear contour plot of local density of
state. The impurity induced bound state is clearly visible near $x=0$
and $\omega=0$. }

\label{fig4} 
\end{figure}

In Fig. 4, we show the local density of state $\rho(x,\omega)$. The
mid-gap bound state can be easily identified in spatially resolved
rf spectroscopy, which is a cold-atom analog of scanning tunneling
microscopy (STM). If such a bound state exists, one would observe
a strong rf-signal at around origin and zero energy, which decays
exponentially in space and energy. The maximum rf-signal, however,
is located slightly away from the origin, as the total density is
completely depleted right at the impurity site.

The existence of a mid-gap state in the topological superfluid phase
is certainly not consistent with Anderson's theorem \cite{Anderson1959}
for potential scattering in \textit{s}-wave superconductors. However,
it can be understood from the combined effect of the spin-orbit coupling
and effective Zeeman field. Beyond the critical Zeeman field $h_{c}$,
the Fermi cloud is actually a \textit{p}-wave-like superfluid (see,
for example, the discussion in Sec. IIA of Ref. \cite{Wei2012}).
This is also the underlying reason why the cloud is in a topological
state. For superfluids with a non-zero angular momentum order parameter,
non-magnetic impurity is a pair-breaker and would lead to a mid-gap
bound state.

\subsection{Universal mid-gap state}

An impurity-induced bound state is not a unique feature of topological
superfluids, as it can also exist in superfluids with even-parity
angular momentum order parameter, such as \textit{d}-wave and \textit{g}-wave
superfluids. Here, however, we argue that the existence of a deep,
universal in-gap bound state in the limit of strong impurity scattering
would be a robust feature of topological superfluids. Despite of the
details of impurity scattering (i.e., non-magnetic or magnetic impurity,
positive or attractive scattering potential), we would observe exactly
the same bound state, when the impurity scattering strength is strong
enough. This argument is based on the consideration that a strong
impurity will always deplete the atoms at the impurity site and hence
create a vacuum area that is topologically trivial. Thus, at the interface
between the topologically non-trivial and trivial areas, we would
observe a pair of Majorana edge states \cite{Sau2012} - the precursor
of the universal bound state. Ideally, the energy of the universal
bound state will be zero.

\begin{figure}[htp]
\begin{centering}
\includegraphics[clip,width=0.48\textwidth]{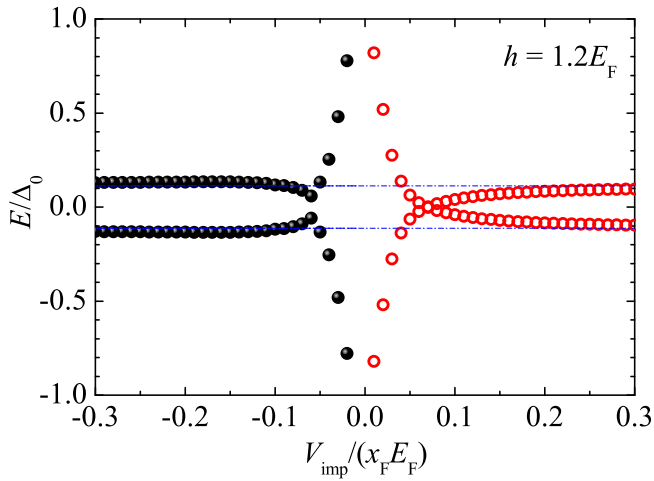} 
\par\end{centering}

\caption{(color online) The dependence of the bound state energy on the impurity
strength for a topological superfluid at $h=1.2E_{F}$. The solid
and empty circles show the results for attractive and repulsive potential
scattering, respectively. The dashed lines gives the bound state energy
at the infinitely large impurity strength, $E\simeq\pm0.113\Delta_{0}$,
obtained by an extrapolation. Here, $\Delta_{0}\simeq0.464E_{F}$
is the pairing gap at the trap center without impurity.}

\label{fig5} 
\end{figure}

In Fig. 5, we plot the energy of the mid-gap bound state as a function
of the impurity scattering strength at $h=1.2E_{F}$. Indeed, when
the absolute value of the scattering strength $V_{imp}$ is sufficiently
large, the energy of the bound state converges to a single value,
$E\simeq0.113\Delta_{0}$, where $\Delta_{0}\simeq0.464E_{F}$ is
the pairing gap at the trap center in the absence of impurity (see
Fig. 3). We have also checked the case with a magnetic impurity and
have found the same bound state energy (not shown in the figure).
The same bound state energy, found under different type of strong
impurities, is a clear indication of the emergence of a universal
impurity-induced bound state. It would also be a unique feature of
the existence of a topological superfluid.

\begin{figure}[htp]
\begin{centering}
\includegraphics[clip,width=0.48\textwidth]{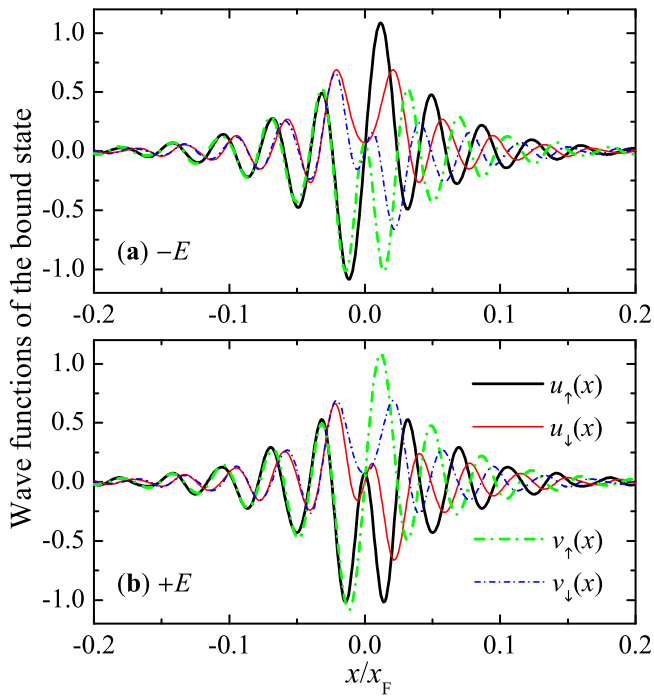} 
\par\end{centering}

\caption{(color online) Wave-function of the universal bound state with energy
$E\simeq\pm0.113\Delta_{0}$, for a topological superfluid at $h=1.2E_{F}$.
The wave-function at $\pm E$ may be regarded as the bond and anti-bond
superposition of two Majorana wave-functions, which satisfy respectively
the symmetry $u_{\sigma}\left(x\right)=\nu_{\sigma}^{*}\left(x\right)$
(on the left side with $x<0$) and $u_{\sigma}\left(x\right)=-\nu_{\sigma}^{*}\left(x\right)$
(on the right side with $x>0$). }

\label{fig6} 
\end{figure}

We note, however, that the bound state energy is not precisely zero
as we may anticipate from the Majorana edge-state picture as mentioned
in the above. This is due to the fact that a pair of zero-energy Majorana
fermions, localized at the same position (i.e., impurity site), could
interfere with each other, leading to a small energy splitting whose
magnitude would depend on the detailed configuration of the Fermi
cloud. In Fig. 6, we present the wave-function of the universal impurity-induced
bound state. Indeed, the wave-function of the universal bound state
can be viewed as the bond and anti-bond superposition of the wave-functions
of two Majorana fermions, which satisfy the symmetry of $u_{\sigma}\left(x\right)=\nu_{\sigma}^{*}\left(x\right)$
or $u_{\sigma}\left(x\right)=-\nu_{\sigma}^{*}\left(x\right)$, respectively.

We note also that the mid-gap state induced by non-magnetic impurities
in topological superconducting nanowires has recently been predicted
by Sau and Demler, based on a non-self-consistent \textit{T}-matrix
and Green function method \cite{Sau2012}. By increasing the impurity
strength, it was reported that the bound state energy saturates to
zero-energy, instead of converging to a nonzero value. In addition,
a shallow bound state was predicted in the non-topological superconducting
phase with spin-orbit coupling. These predictions are different from
our numerical results. We ascribe these discrepancies to the lack
of self-consistency in the \textit{T}-matrix approach.

\subsection{Realistic gaussian-shape impurity}

We consider so far a delta-like impurity scattering potential. In
real experiments, the non-magnetic impurity would be simulated by
an off-resonant dimple laser light, which has a finite width in space.
Thus, it is more reasonable to simulate the impurity by using a gaussian-shape
scattering potential.

\begin{figure}[htp]
\begin{centering}
\includegraphics[clip,width=0.48\textwidth]{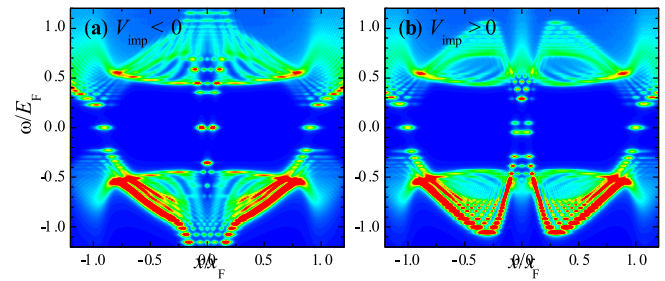} 
\par\end{centering}

\caption{(color online) Linear contour plot of density of state at $h=1.2E_{F}$,
for an attractive or a repulsive gaussian-shape impurity scattering
potential. Here, we take $d=0.1x_{F}$ and $V_{imp}=\pm0.30x_{F}E_{F}$. }

\label{fig7} 
\end{figure}

Fig. 7 reports the linear contour plot of local density of state at
$h=1.2E_{F}$ for a strong attractive (a) and repulsive (b) gaussian-shape
impurity potential. With a finite width $d=0.1x_{F}$, we observe
a series of bound states in the vicinity of the impurity site. The
lowest-energy bound state is close to the universal bound state that
we find earlier with a delta-like impurity potential.

To give some realistic parameters, let us consider a spin-orbit coupled
Fermi gas of $^{40}$K atoms confined to a tight 2D optical lattice,
with an axial trapping frequency $\omega=2\pi\times116$ Hz \cite{Wang2012}.
By assuming the number of atoms $N=100$ in each tube, the Fermi energy
or temperature is about $300$ nK. We may take $k_{F}\simeq2k_{R}$
and a Raman strength $\Omega_{R}\simeq10E_{R}$, where $E_{R}$ is
the recoil energy. We may anticipate a topological superfluid at temperature
$T<10$ nK. The typical size of the Fermi cloud is about $x_{F}\simeq15$
$\mu m$. Thus, we may use an off-resonant dimple laser with width
$d\simeq1.5$ $\mu m$ to simulate the non-magnetic impurity. The
strength of the impurity can be easily tuned by controlling the strength
of the dimple laser light. With these parameters, we may be able to
observe the universal impurity-induced bound state discussed in the
above.

\section{Conclusions}

In summary, we have argued that a strong non-magnetic impurity will
induce a universal bound state in topological superfluids. This provides
a unique feature to characterize the long-sought topological superfluids.
We have proposed a realistic setup to observe such a universal impurity-induced
bound state in atomic topological superfluids, which are to be realized
in spin-orbit coupled Fermi gases of $^{40}$K atoms. The necessary
conditions, including the realization of spin-orbit coupling by two-photon
Raman process, the achievement of one-dimensional confinement by optical
lattice, and the simulation of non-magnetic impurities using off-resonant
dimple laser light, are all within the current experimental reach.
Therefore, we anticipate our proposal will be realized soon at Shanxi
University in China \cite{Wang2012} or elsewhere.

\section*{Acknowledgments}

We thank Hui Hu for many helpful discussions. This work was supported
by the ARC Discovery Project (Grant No. DP0984637) and the NFRP-China
(Grant No. 2011CB921502).

\appendix
%dummy comment inserted by tex2lyx to ensure that this paragraph is not empty

\section{Solving the BdG equation in one dimension}

We solve the BdG equation Eq. (\ref{bdgeq}) by expanding the Bogoliugbov
wavefunctions $u_{\sigma}(x)$ and $\nu_{\sigma}(x)$ in the basis
of 1D harmonic oscillators $\phi_{j}(x)=(1/\sqrt{\pi^{1/2}2^{j}j!})H_{j}(x)e^{-x^{2}/2}$,
\begin{eqnarray}
u_{\sigma}\left(x\right) & = & \sum_{j=0}^{M-1}U_{\sigma j}\phi_{j}\left(x\right),\\
\nu_{\sigma}\left(x\right) & = & \sum_{j=0}^{M-1}V_{\sigma j}\phi_{j}\left(x\right),
\end{eqnarray}
 Here, $H_{j}(x)$ is the $j$-th Hermite polynomial and, for convenience,
we have used the natural unit in harmonic traps, $m=\hbar=\omega=1$,
so that the oscillator length $a\equiv\sqrt{\hbar/(m\omega)}=1$ and
the oscillator energy $\hbar\omega=1$. On such a basis, the BdG Hamiltonian
Eq. (\ref{BdG_Hami}) is converted to a $4M\times4M$ secular matrix,\begin{widetext}
\begin{equation}
{\cal H}_{BdG}=\left[\begin{array}{cccc}
{\cal H}_{S}^{ij}-h\delta_{ij} & -R_{ij} & 0 & -\Delta_{ij}\\
R_{ij} & {\cal H}_{S}^{ij}+h\delta_{ij} & \Delta_{ij} & 0\\
0 & \Delta_{ij} & -{\cal H}_{S}^{ij}+h\delta_{ij} & R_{ij}\\
-\Delta_{ij} & 0 & -R_{ij} & -{\cal H}_{S}^{ij}-h\delta_{ij}
\end{array}\right],\label{BdG_matrix}
\end{equation}
\end{widetext} where the matrix elements, 
\begin{eqnarray}
{\cal H}_{S}^{ij} & = & \left(i+1/2-\mu\right)\delta_{ij}+V_{imp}^{ij},\\
R_{ij} & = & \lambda\left[\sqrt{j/2}\delta_{i,j-1}-\sqrt{\left(j+1\right)/2}\delta_{i,j+1}\right].
\end{eqnarray}
 To calculate efficiently the matrix elements $V_{imp}^{ij}\equiv\int_{-\infty}^{+\infty}dx\phi_{i}(x)V_{imp}(x)\phi_{j}(x)$
and $\Delta_{ij}\equiv\int_{-\infty}^{+\infty}dx\phi_{i}(x)\Delta(x)\phi_{j}(x)$,
we discretize space $(-L/2,L/2)$ into $N_{grid}$ equally spaced
points, where the simulation length $L$ and the number of grid $N_{grid}$
should be sufficiently large so that the basis function $\phi_{j}(x)$
($j=0,..,M-1$) can be accurately sampled. At the number of atoms
$N=100$, typically we take $M=500$, $N_{grid}=6400$, and $L=70\sqrt{\hbar/(m\omega)}$.
The gaussian impurity potential $V_{imp}\left(x\right)$ and pairing
gap function $\Delta\left(x\right)$, as well as the total density
$n(x)$, will be stored as an array of length $N_{grid}$. We note
that, for a delta-like impurity we immediately have $V_{imp}^{ij}=V_{imp}\phi_{i}\left(x_{0}\right)\phi_{j}\left(x_{0}\right)$.
By diagonalizing the $4M\times4M$ secular matrix Eq. (\ref{BdG_matrix}),
we obtain the quasiparticle energy $E_{\eta}$ and the eigenvector
$U_{\sigma j}^{\eta}$ and $V_{\sigma j}^{\eta}$ ($j=0,..,M-1$).
The latter gives the quasiparticle wave-function $u_{\sigma\eta}\left(x\right)$
and $\nu_{\sigma\eta}\left(x\right)$. Note that, the eigenvector
$U_{\sigma j}^{\eta}$ and $V_{\sigma j}^{\eta}$ have to satisfy
the condition $\sum_{\sigma j}[(U_{\sigma j}^{\eta})^{2}+(V_{\sigma j}^{\eta})^{2}]=1$,
due to the normalization of the quasiparticle wavefunctions, i.e.,
$\int_{-\infty}^{+\infty}dx\sum_{\sigma}[u_{\sigma\eta}^{2}(x)+v_{\sigma\eta}^{2}(x)]=1$.

In the practical calculation, due to computational limitation, we
have to use a finite expansion basis. This is controlled by the cut-off
$M$ for the number of 1D harmonic oscillators. Furthermore, we must
impose a high energy cut-off $E_{c}$ for the quasiparticle energy
levels. To make our result cut-off independent, we adopt a hybrid
approach, in which we solve the discrete BdG equation for the energy
levels below the high energy cut-off $E_{c}$. While above $E_{c}$,
we use a semiclassical plane-wave approximation for the wavefunctions
that should work very well for high-lying energy levels. For simplicity,
to take the semiclassical approximation we may neglect the spin-orbit
coupling term $R_{ij}$ in the BdG Hamiltonian Eq. (\ref{BdG_matrix}).
In the end, for the pairing gap function and the total density, we
shall use the semiclassical expressions listed in the Sec. IVC of
Ref. \cite{Liu2007}. To summarize briefly, the contributions of discrete
low-lying energy levels (labeled by an index ``$\eta$'') and continuous
high-lying energy levels to the total density are given by,\begin{widetext}
\begin{equation}
n_{d}\left(x\right)=\frac{1}{2}\sum_{\left|E_{\eta}\right|<E_{c}}\sum_{\sigma}\left[\left|u_{\sigma\eta}\left(x\right)\right|^{2}f\left(E_{\eta}\right)+\left|v_{\sigma\eta}\left(x\right)\right|^{2}f\left(-E_{\eta}\right)\right]
\end{equation}
 and 
\begin{equation}
n_{c}\left(x\right)=\frac{\sqrt{2m}}{4\pi\hbar}\left(\int\limits _{E_{c}+h}^{+\infty}+\int\limits _{E_{c}-h}^{+\infty}\right)d\epsilon\frac{\left[\epsilon/\sqrt{\epsilon^{2}-\Delta^{2}\left(x\right)}-1\right]}{\sqrt{\mu+\sqrt{\epsilon^{2}-\Delta^{2}\left(x\right)}}},
\end{equation}
 respectively. For the pairing gap function, we have 
\begin{equation}
\Delta(x)=-\frac{g_{1D}^{eff}\left(x\right)}{2}\sum_{\left|E_{\eta}\right|<E_{c}}\sum_{\sigma}\left[u_{\uparrow\eta}\left(x\right)v_{\downarrow\eta}^{*}\left(x\right)f\left(E_{\eta}\right)+u_{\downarrow\eta}\left(x\right)v_{\uparrow\eta}^{*}\left(x\right)f\left(-E_{\eta}\right)\right],\label{gapeff}
\end{equation}
\end{widetext} where the effective interaction strength $g_{1D}^{eff}\left(x\right)$
is determined by, 
\begin{equation}
\frac{1}{g_{1D}^{eff}\left(x\right)}=\frac{1}{g_{1D}}+\frac{\sqrt{2m}}{4\pi\hbar}\int\limits _{E_{c}-h}^{\infty}d\epsilon\frac{1}{\sqrt{\epsilon^{2}-\Delta^{2}\left(x\right)}}.\label{g1deff}
\end{equation}

The numerical procedure of solving the BdG equation is therefore as
follows. For a given set of parameters ($N$, $g_{1D}$, $h$, $\lambda$,
and $T$), we (1) start with an initial guess or a previously determined
better estimate for $\Delta\left(x\right)$, (2) solve Eq. (\ref{g1deff})
for the effective coupling constant, (3) then solve Eq. (\ref{BdG_matrix})
for all the quasiparticle wavefunctions up to the chosen energy cut-off
to find $u_{\sigma\eta}\left(x\right)$ and $v_{\sigma\eta}\left(x\right)$,
and finally determine an improved value for the order parameter from
Eq. (\ref{gapeff}). During the iteration, the total density $n(x)=n_{d}(x)+n_{c}(x)$
is updated. The chemical potentials $\mu$ is adjusted slightly in
each iterative step to enforce the number-conservation condition $\int_{-\infty}^{+\infty}dxn(x){\bf =}N$,
until final convergence is reached.

\subsection{Check on the numerical accuracy}

We have checked carefully the numerical accuracy of our hybrid approach
at different sets of parameters and at both zero temperature and finite
temperatures. In Fig. 8, we check the dependence on the cut-off energy
$E_{c}$ at $h=1.2E_{F}$ in the absence of impurity scattering. The
pairing gap function becomes essentially independent on $E_{c}$ once
$E_{c}\geq6E_{F}$, with a relative error less than $1$\%. The cut-off
energy dependence for the total density is even weaker (not shown
in the figure). Thus, we conclude that our hybrid calculation is quantitatively
reliable with $E_{c}=6E_{F}$. At this energy cut-off, each iteration
in the self-consistent calculation takes approximately several minutes,
by using a standard desktop computer. The convergence for a set of
parameters is typically reached after $20-50$ iterations.

\begin{figure}[htp]
\begin{centering}
\includegraphics[clip,width=0.48\textwidth]{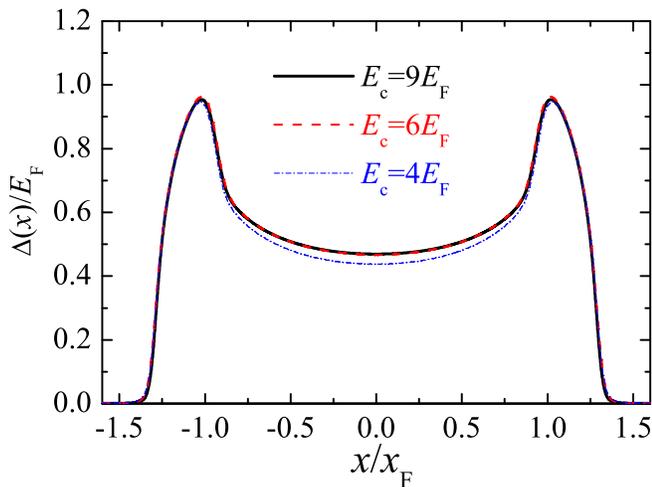} 
\par\end{centering}

\caption{(color online) The dependence of the pairing gap function on the high-energy
cut-off $E_{c}$. Here, we consider a spin-orbit coupled Fermi gas
of $N=100$ atoms in harmonic traps at zero temperature. The dimensionless
interaction parameter is $\gamma=\pi$. The spin-orbit coupling is
taken to be $\lambda k_{F}/E_{F}=1$ and $h=1.2E_{F}$. At these typical
parameters, the results become independent on the cut-off energy $E_{c}$
once it is larger than $6E_{F}$. }

\label{fig8} 
\end{figure}


\begin{thebibliography}{10}
\bibitem{Balatsky2006} A. V. Balatsky, I. Vekhter, and J.-X. Zhu,
Rev. Mod. Phys. \textbf{78}, 373 (2006).

\bibitem{Mackenzie1998} A. P. Mackenzie, R. K. W. Haselwimmer, A.
W. Tyler, G. G. Lonzarich, Y. Mori, S. Nishizaki, and Y. Maeno, Phys.
Rev. Lett. \textbf{80}, 161 (1998).

\bibitem{Millis2003} A. J. Millis, Solid State Commun. \textbf{126},
3 (2003).

\bibitem{Qi2011} X.-L. Qi and S.-C. Zhang, Rev. Mod. Phys. \textbf{83},
1057 (2011).

\bibitem{Majorana1937} E. Majorana, Nuovo Cimennto \textbf{14}, 171
(1937).

\bibitem{Wilczek2009} F. Wilczek, Nat. Phys. \textbf{5}, 614 (2009).

\bibitem{Kitaev2006} A. Kitaev, Ann. Phys. (NY) \textbf{321}, 2 (2006).

\bibitem{Nayak2008} C. Nayak, S. Simon, A. Stern, M. Freedman, and
S. Das Sarma, Rev. Mod. Phys. \textbf{80}, 1083 (2008).

\bibitem{Mourik2012} V. Mourik, K. Zuo, S. M. Frolov, S. R. Plissard,
E. P. A. M. Bakkers, and L. P. Kouwenhoven, Science \textbf{336},
1003 (2012).

\bibitem{Rokhinson2012} L. P. Rokhinson, X. Liu, and J. K. Furdyna,
arXiv:1204.4214 (2012).

\bibitem{Das2012} A. Das, Y. Ronen, Y. Most, Y. Oreg, M. Heiblum,
and H. Shtrikman, arXiv:1205.7073 (2012).

\bibitem{Fu2008} L. Fu and C. L. Kane, Phys. Rev. Lett. \textbf{100},
096407 (2008).

\bibitem{Alicea2011} J. Alicea, Y. Oreg, G. Refael, F. von Oppen
and M. P. A. Fisher, Nature Phys. \textbf{7}, 412 (2011).

\bibitem{Jiang2011} L. Jiang, T. Kitagawa, J. Alicea, A. R. Akhmerov,
D. Pekker, G. Refael, J. I. Cirac, E. Demler, M. D. Lukin, and P.
Zoller, Phys. Rev. Lett. \textbf{106}, 220402 (2011).

\bibitem{Liu2012} X.-J. Liu and H. Hu, Phys. Rev. A \textbf{85},
033622 (2012).

\bibitem{Wei2012} R. Wei and E. J. Mueller, arXiv:1208.5450 (2012).

\bibitem{Bloch2008} I. Bloch, J. Dalibard, and W. Zwerger, Rev. Mod.
Phys. \textbf{80}, 885 (2008).

\bibitem{Chin2010} C. Chin, R. Grimm, P. Julienne, and E. Tiesinga,
Rev. Mod. Phys. \textbf{82}, 1225 (2010).

\bibitem{Hu2007} H. Hu, X.-J. Liu, and P. D. Drummond, Phys. Rev.
Lett. \textbf{98}, 070403 (2007).

\bibitem{Liao2010} Y. A. Liao, A. S. C. Rittner, T. Paprotta, W.
Li, G. B. Partridge, R. G. Hulet, S. K. Baur, and E. J. Mueller, Nature
(London) \textbf{467}, 567 (2010).

\bibitem{Wang2012} P. Wang, Z.-Q. Yu, Z. Fu, J. Miao, L. Huang, S.
Chai, H. Zhai and J. Zhang, Phys. Rev. Lett. \textbf{109}, 095301
(2012).

\bibitem{Cheuk2012} L. W. Cheuk, A. T. Sommer, Z. Hadzibabic, T.
Yefsah, W. S. Bakr, and M. W. Zwierlein, Phys. Rev. Lett. \textbf{109},
095302 (2012).

\bibitem{OurImpurity} L. Jiang, L. O. Baksmaty, H. Hu, Y. Chen, and
H. Pu, Phys. Rev. A \textbf{83}, 061604(R) (2011).

\bibitem{Palencia2010} L. Sanchez-Palencia and M. Lewenstein, Nature
Phys. \textbf{6}, 87 (2010).

\bibitem{Sau2012} J. D. Sau and E. Demler, arXiv:1204.2537 (2012).

\bibitem{Yan2012} Z. Yan, X. Yang, L. Sun, and S. Wan, arXiv:1204.0571
(2012).

\bibitem{Bergeman2003} T. Bergeman, M. G. Moore, and M. Olshanii,
Phys. Rev. Lett. \textbf{91}, 163201 (2003).

\bibitem{Liu2007} X.-J. Liu, H. Hu, and P. D. Drummond, Phys. Rev.
A \textbf{76}, 043605 (2007).

\bibitem{Liu2008} X.-J. Liu, H. Hu, and P. D. Drummond, Phys. Rev.
A \textbf{78}, 023601 (2008).

\bibitem{Lutchyn2010} R. M. Lutchyn, J. D. Sau, and S. D. Sarma,
Phys. Rev. Lett. \textbf{105}, 077001 (2010).

\bibitem{Oreg2010} Y. Oreg. G. Refael, and F. von Oppen, Phys. Rev.
Lett. \textbf{105}, 177002 (2010).

\bibitem{Anderson1959} P. W. Anderson, J. Phys. Chem. Solids \textbf{11},
26 (1959).\end{thebibliography}
\end{document}